# The evolution of tyrosine-recombinase elements in Nematoda

Amir Szitenberg[1], Georgios Koutsovoulos[2], Mark L Blaxter[2] and David H Lunt[1]

[1]Evolutionary Biology Group, School of Biological, Biomedical & Environmental Sciences, University of Hull, Hull, HU6 7RX, UK

[2]Institute of Evolutionary Biology, The University of Edinburgh, EH9 3JT, UK

A.Szitenberg@hull.ac.uk

## Abstract

Transposable elements can be categorised into DNA and RNA elements based on their mechanism of transposition. Tyrosine recombinase elements (YREs) are relatively rare and poorly understood, despite sharing characteristics with both DNA and RNA elements. Previously, the Nematoda have been reported to have a substantially different diversity of YREs compared to other animal phyla: the *Dirs1*-like YRE retrotransposon was encountered in most animal phyla but not in Nematoda, and a unique *Pat1*-like YRE retrotransposon has only been recorded from Nematoda. We explored the diversity of YREs in Nematoda by sampling broadly across the phylum and including 34 genomes representing the three classes within Nematoda. We developed a method to isolate and classify YREs based on both feature organization and phylogenetic relationships in an open and reproducible workflow. We also ensured that our phylogenetic approach to YRE classification identified truncated and degenerate elements, informatively increasing the number of elements sampled. We identified *Dirs1*-like elements (thought to be absent from Nematoda) in the nematode classes Enoplia and Dorylaimia indicating that nematode model species do not adequately represent the diversity of transposable elements in the phylum. Nematode *Pat1*-like elements were found to be a derived form of another *Pat1*-like element that is present more widely in animals. Several sequence features used widely for the classification of YREs were found to be homoplasious, highlighting the need for a phylogenetically-based classification scheme. Nematode model species do not represent the diversity of transposable elements in the phylum.

## Introduction

### Transposable elements

Transposable elements (TE) are mobile genetic elements capable of propagating within a genome and potentially transferring horizontally between organisms (Nakayashiki 2011). They typically constitute significant proportions of bilaterian genomes, comprising 45% of the human genome (Lander et al. 2001), 22% of the *Drosophila melanogaster* genome (Kapitonov and Jurka 2003) and 12% of the *Caenorhabditis elegans* genome (Bessereau 2006). TEs may also have important evolutionary effects, such as promoting alternative splicing (Sorek et al. 2002), inducing variation accumulation under stress (Badyaev 2005) and increasing the genetic load (Wessler 2006).

TEs can be broadly divided into DNA and RNA classes. DNA TEs (transposons) transfer as dsDNA, leaving a vacant locus at the point of origin, together with a target site duplication (TSD) (Wessler 2006). They are thought to increase in copy number via various recombination related mechanisms between vacant and populated TE loci, partly due to the similarity of TSDs across the genome (Wessler, 2006). RNA TEs (retrotransposons and retroposons) do not exit their locus of origin but rather propagate through the reverse transcription of an RNA intermediate copy back into an additional site in the genome (Finnegan 2012). RNA elements are usually the most numerous type of TE and can have tens of thousands, or even millions, of copies in a single genome (Finnegan 2012). Despite this there can be variation in the relative proportions and some species have DNA elements as the most frequent class, as the case is in *C. elegans* (*Caenorhabditis elegans* Sequencing Consortium 1998).

**Tyrosine recombinase TEs**

Tyrosine Recombinase Elements (YREs) are found in both the DNA and RNA TE classes. They contain a Tyrosine Recombinase (YR) domain that replaces the transposase and integrase proteins encoded in DNA and RNA TEs, respectively. The YR domain facilitates transposition without forming a TSD. YREs have been suggested to have emerged from a single or several events of recombination between DNA and RNA elements (Kojima and Jurka 2011) which makes them interesting an important group for understanding the evolution and maintenance of TEs more generally. YREs are diverse in sequence and structure, but this diversity is not equally represented across the animal phyla (Poulter and Goodwin 2005; Kojima and Jurka 2011; Piednoël et al. 2011), and their evolutionary history can sometimes be puzzling. Nematoda for example are known to have one unique form of YREs (a form of *Pat1* found only in this phylum) and to entirely lack another (*Dirs1)*, which is otherwise relatively common (Piednoël et al. 2011). However, the diversity of YREs in Nematoda is still poorly understood and a phylogenetically informed analysis with broad taxonomic sampling of both the YREs and their hosts is required to thoroughly address the subject.

**YRE classification**

DNA YREs possess only a YR protein domain and include the Crypton and TEC elements. Although Cryptons were first discovered in fungi (Goodwin et al. 2003), four distinct, possibly polyphyletic, lineages have been defined in fungi, diatoms and animals (Kojima and Jurka 2011). It is thought that Cryptons may have contributed to the origin of RNA YREs (Kojima and Jurka 2011). TEC elements, by contrast, appear to have a very limited taxonomic distribution and are currently known only from ciliates (Doak et al. 1994; Jacobs et al. 2003).

RNA YREs, like other long terminal repeat (LTR) retrotransposons, possess the capsid protein Gag, and a polyprotein that includes the reverse transcriptase (RT) and RNase H (RH) domains. LTR retrotransposons (Gypsy, Copia and Bell) may have been the source of the ancestral RNA element of the YRE ancestor (Kojima and Jurka 2011). Unlike the LTR retrotransposons, YRE retrotransposons possess the YR domain and lack the integrase gene (Poulter and Goodwin 2005; Wicker et al. 2007). They sometimes also encode a methyltransferase (MT) domain.

**Structure based classification of YREs**

A set of molecular sequence features are widely used to classify YRE retrotransposons: the presence and strand of the RT, MT and YR domains, the presence of a zinc-finger (ZF) motif in the Gag protein, and the presence and relative arrangement of characteristic repeat sequences (Cappello et al. 1984; Cappello et al. 1984; Goodwin and Poulter 2001; Goodwin et al. 2004; Piednoel and Bonnivard 2009; Piednoël et al. 2011; Muszewska et al. 2013; fig. 1). Three groups of YRE retrotransposons have been defined: DIRS, Ngaro and Viper (Goodwin and Poulter 2004; Lorenzi et al. 2006).

DIRS, in turn, comprise *Dirs1*-like elements and PAT elements, and PAT can be broken down again to include *Pat1*-like elements, *Toc* elements and *Kangaroo* elements (fig. 1).

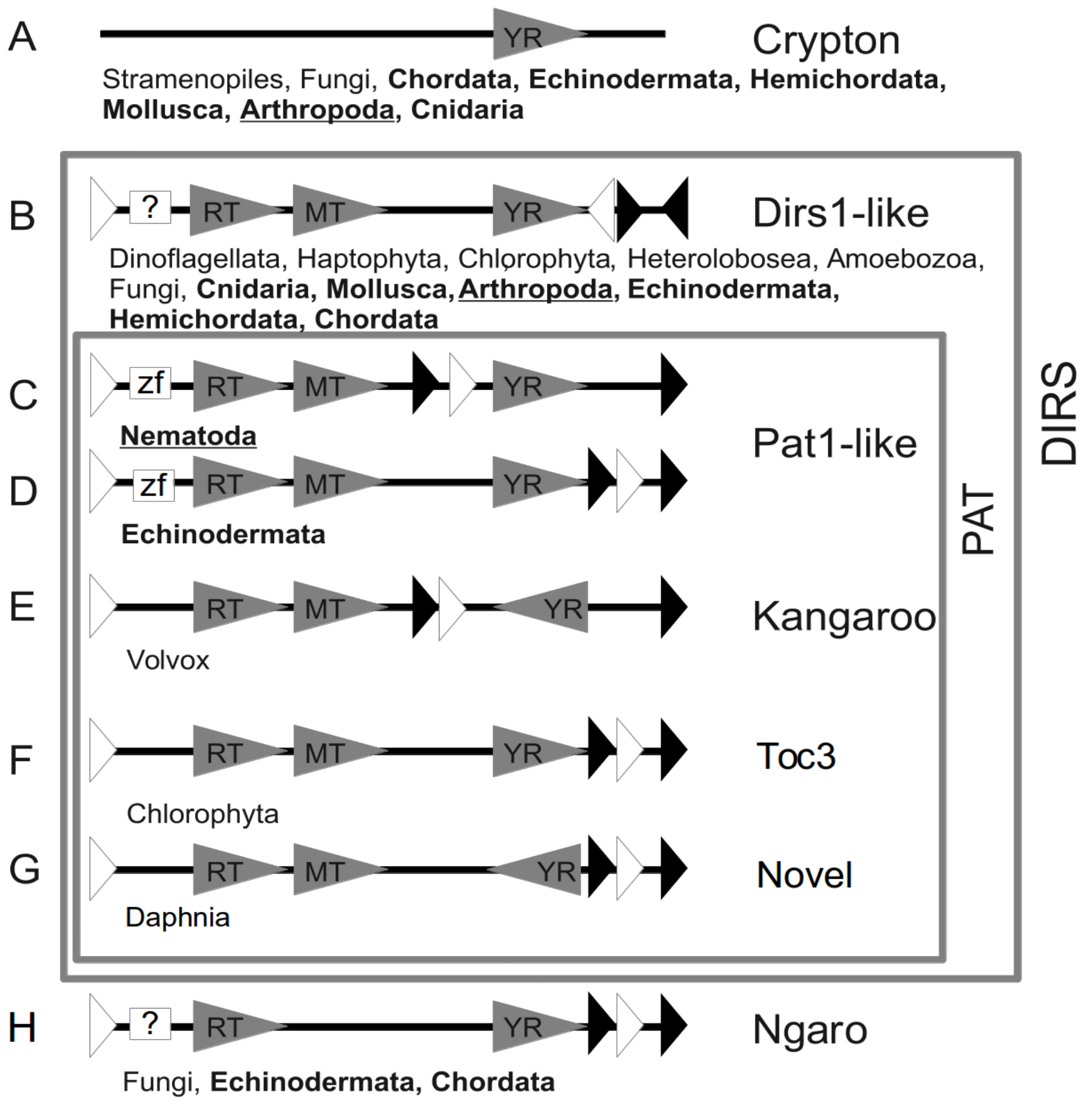

**Fig. 1: The diversity of tyrosine recombinase elements (YREs) and their diagnostic features for taxonomic classification** The known taxonomic distribution of each element (A - H) is listed along with a cartoon of its structure. Metazoa are in bold font and Ecdysozoa are underlined. The features considered are the presence and absence of the reverse transcriptase (RT), methyltransferase (MT) and tyrosine recombinase (YR) domains and their direction (grey triangles), as well as the presence, absence and position of split direct repeats (pairs of triangles, sharing a colour and pointing in the same direction), inverted repeats (pairs of triangles, sharing a colour and pointing in opposite directions) and zinc-finger motifs (zf) from the Gag protein. Where a question mark is indicated, some members of the group possess and others lack a zf motif.

DIRS elements are YRE retrotransposons that encode a putative MT domain. Within the DIRS group, *Dirs1*-like elements and PAT elements are differentiated by the presence of two consecutive pairs of inverted repeats in *Dirs1*-like (fig. 1B) and split direct repeats in PAT elements (fig. 1C-1F). *Dirs1*-like elements were discovered in Amoebozoa (Cappello et al. 1984) and are also present in Viridiplantae, Metazoa and other eukaryotes. (Piednoël et al. 2011). Like other YRE retrotransposons they have internal repeats that couple with the terminal ones (fig. 1B). For a detailed description of *Dirs1* repeat sequences see Piednoël et al. (2011).

PAT elements (fig. 1C-1F) differ from *Dirs1*-like elements by the presence of direct-split repeats. The repeats are also referred to as A1-B1-A2-B2 repeats where A2 is an identical repeat of A1 and B2 of B1. As mentioned, the PAT group includes *Pat1*-like, *Kangaroo* and *Toc* elements. *Pat1* elements (fig. 1C-1D) were first identified in the nematode *Panagrellus redivivus* (Panagrolaimomorpha; Tylenchina; Rhabditida) (de Chastonay et al. 1992) (see figure 2 for relationships of species analysed) and subsequently also in the nematodes *Caenorhabditis briggsae* (Rhabditomorpha; Rhabditina; Rhabditida) (Goodwin and Poulter 2004) and *Pristionchus pacificus* (Diplogasteromorpha; Rhabditina; Rhabditida) (Piednoël et al. 2011). A distinct form of *Pat1*-like elements was described from the sea urchin *Strongylocentrotus purpuratus* (Echinodermata) (Goodwin and Poulter 2004). The Nematoda-form *Pat1*-like elements (fig. 1C) differs from the echinoderm-form (fig. 1D) in the placement of their internal repeat (A2B1) sequences. Both forms have a zinc-finger motif in the Gag protein, which is absent from the other PAT elements, *Kangaroo* and *Toc*. The *Kanagaroo* element, found in *Volvox carteri* (Chlorophyta; Viridiplantae) (Duncan et al. 2002), differs from other PAT elements by having an inverted YR domain (fig. 1E) and by the absence of a zinc-finger motif. In *Kangaroo* elements, the internal repeats are located between the MT and YR domains (as observed in the Nematoda-form *Pat1*-like elements). *Toc3* PAT elements (fig. 1F) were found in algae (Goodwin and Poulter 2004) and differ from *Pat1*-like elements by the absence of a zinc-finger motif, and from *Kangaroo* elements by the direction of the YR domain.

Ngaro and Viper are two groups of non-DIRS YRE retrotransposons. These predominantly differ from DIRS elements by the absence of the putative MT domain (fig 1G). Like PAT elements they possess split direct repeats, with the internal repeats found downstream to the YR domain (Goodwin and Poulter 2004; Wicker et al. 2007). Ngaro elements were originally found in *Danio rerio* (zebrafish; Osteichthyes; Chordata), *S. purpuratus* and fungi (Poulter and Goodwin 2005), while Viper elements are found in *Trypanosoma* (Trypanosomatidae; Kinetoplastida) (Lorenzi et al. 2006).

In spite of their exceptional diversity, YREs are quite rare. *Dirs1* from *Dictyostelium discoideum* (Amoebozoa; Cappello et al. 1984) is present in 40 intact copies and 200 - 300 fragments. Crypton (fig. 1A) is present in a few dozen copies in each of a range of eukaryote species (Kojima and Jurka 2011). TEs with such small population size, however, will be subject to strong genetic drift and variation in copy number, and thus will be prone to elimination (Collins et al. 1987). Nematoda are considered to have undergone a shift in their YRE content compared to other phyla, losing *Dirs1*-like elements (fig. 1B) and expanding *Pat1*-like elements (Piednoël et al. 2011). However, the true diversity of YREs in Nematoda in not known as current estimates are based largely on a few, relatively closely related species (*P. redivivus*, *P. pacificus* and *C. elegans*). Here we survey whole genome sequencing data from a wide taxonomic range of nematode species and show that a shift in YRE content has indeed occurred. However, *Dirs1*-like elements are present in at least one of the three Nematoda classes, and the Nematoda form of *Pat1*-like elements is closely related to *Pat1* elements from other animal phyla.

## New Approaches

To identify and quantify YREs in nematodes, we utilized homology based search methods to locate YREs, made a preliminary classification based on characteristic features, and used phylogenetic methods to refine and corroborate these classifications. We conducted further phylogenetic analyses to classify partial or degenerate elements relative to complete elements. This stage allowed us to include partial and potentially degraded elements in the copy number counts and have a better understanding of the origins of the distribution of YREs among the nematode species. Unlike similarity based clustering methods (e.g., Piednoël et al. 2011; Muszewska et al. 2013; Guérillot et al. 2014; Iyer et al. 2014), a phylogenetic approach accounts for homoplasy and is better adapted for the analysis of potentially degraded sequences. A diagram showing the stages of our analysis is in figure S1. In order to facilitate replication and extension of our work with new genomic data we have made all our analysis steps reproducible through use of an iPython notebook and github repository that include all analysis code and intermediate data sets (http://dx.doi.org/10.6084/m9.figshare.1004150).

## Results

We identified putative homologues of three YRE protein domains, YR, RT and MT, in genome assemblies of 34 nematode and 12 outgroup species. Over 2,500 significant matches to YREs were found in 24 species (table 1). These were first classified based on the presence, absence and direction of YRE sequence features (fig. 1). Although only 207 elements in 13 of the assemblies could be classified unequivocally based on these diagnostic features, these classified sequences were useful additional reference sequences, complementing the ones obtained from Retrobase (http://biocadmin.otago.ac.nz/fmi/xsl/retrobase/home.xsl) and Repbase-update (Jurka et al. 2005) (fig. 2). In addition, we used them to corroborate the results of subsequent phylogenetic analyses.

Our phylogenetic classification, based on YR domain sequences, included two steps. In the first step, only complete elements, for which terminal repeats were identified, were considered, in order to deliniate YRE clades. In the second step, all the putative YR matches were included, in order to classify partial elements based on their phylogenetic relationships with complete elements. After this phylogenetic classification (fig. S2), 963 elements were classified in 17 genomes (fig. 3).

To assess whether the genome assemblies used were of sufficient quality to permit YRE discovery, we also searched for RT domains from three LTR elements, Gypsy, Copia and BELL, reasoning that if we were unable to detect any of the abundant LTR class elements it was likely that the assembly was too poor. The N50 contig lengths of the assemblies (table 1) did not correlate with the number of YRE matches (linear $R^2 = 2*10^{-3}$, power $R^2 = 8*10^{-3}$). A greater number of matches were found in outgroup taxa with larger genomes than Nematoda. No species had zero matches in all four searches (YRE plus three LTR searches). *Litomosoides sigmodontis* had the lowest number of matches, including only three to BEL LTR retrotransposons, while *Oscheius tipulae* had 10 or less matches in any searches. *Bursaphelenchus xylophilus, Caenorhabditis angaria* and *Caenorhabditis* sp. 11 had a maximum of 27 matches in any of the searches. For the remaining species, at least 40 matches were found in at least one of the searches. Given these findings, we are confident that cases where no YREs were found usually indicate a real absence, or extreme scarcity, of YREs in those species.

**Table1:** Mean contig lengths, contig length at N50 and PSIBLASTN putative transposable element counts of the genomes analysed

| Code | Species | Mean contig length | N50 contig length | YRE matches | YRE ID by features | YRE ID by phylogeny | BEL | Copia | Gypsy |
|---|---|---|---|---|---|---|---|---|---|
| Haor | *Howardula aoronymphium* | 411 | 429 | 7 | 0 | 0 | 43 | 4 | 62 |
| Ebre | *Enoplus brevis* | 477 | 506 | 459 | 0 | 9 | 513 | 50 | 535 |
| Ovol | *Onchocerca volvulus* | 1,146 | 1265 | 1 | 0 | 0 | 61 | 0 | 2 |
| Mflo | *Meloidogyne floridensis* | 1,231 | 3516 | 0 | 0 | 0 | 100 | 0 | 90 |
| Cang | *Caenorhabditis angaria* | 3,062 | 79858 | 11 | 0 | 2 | 20 | 0 | 14 |
| Wban | *Wuchereria bancrofti* | 3,149 | 5161 | 0 | 0 | 0 | 65 | 0 | 3 |
| Ooch | *Onchocerca ochengi* | 3,970 | 12317 | 0 | 0 | 0 | 156 | 0 | 1 |
| Otip | *Oscheius tipulae* | 3,998 | 13984 | 12 | 0 | 0 | 10 | 1 | 8 |
| Dsim | *Drosophila simulans* | 4,029 | 7074 | 0 | 0 | 0 | 557 | 114 | 642 |
| Hcon | *Haemonchus contortus* | 4,991 | 13338 | 40 | 7 | 18 | 582 | 0 | 448 |
| Dimm | *Dirofilaria immitis* | 5,498 | 71281 | 1 | 0 | 0 | 46 | 0 | 0 |
| Rcul | *Romanomermis culicivorax* | 5,580 | 20133 | 267 | 6 | 70 | 538 | 9 | 682 |
| Agam | *Anopheles gambiae* | 7,667 | 1505544 | 1 | 0 | 0 | 664 | 263 | 654 |
| Asuu | *Ascaris suum* | 8,420 | 290558 | 1 | 0 | 0 | 62 | 0 | 47 |
| Minc | *Meloidogyne incognita* | 8,607 | 12786 | 0 | 0 | 0 | 157 | 1 | 91 |
| C5sp | *Caenorhabditis* sp. 5 | 8,636 | 25228 | 0 | 0 | 0 | 98 | 0 | 50 |
| Cjap | *Caenorhabditis japonica* | 8,835 | 94149 | 62 | 2 | 50 | 206 | 4 | 261 |
| Tspi | *Trichinella spiralis* | 9,256 | 6373445 | 0 | 0 | 0 | 220 | 29 | 140 |
| Ppac | *Pristionchus pacificus* | 9,539 | 1244534 | 57 | 7 | 33 | 156 | 0 | 124 |
| Dviv | *Dictyocaulus viviparus* | 9,562 | 22560 | 0 | 0 | 0 | 95 | 0 | 3 |
| Bmal | *Brugia malayi* | 9,578 | 191089 | 0 | 0 | 0 | 110 | 0 | 3 |
| Hduj | *Hypsibius dujardini* | 10,223 | 50531 | 58 | 0 | 17 | 50 | 0 | 86 |
| Avit | *Acanthocheilonema viteae* | 11,382 | 25808 | 0 | 0 | 0 | 41 | 0 | 3 |
| Bxyl | *Bursaphelenchus xylophilus* | 13,490 | 949830 | 0 | 0 | 0 | 21 | 0 | 14 |
| Hbac | *Heterorhabditis bacteriophora* | 14,630 | 33765 | 2 | 0 | 0 | 45 | 0 | 41 |
| Mhap | *Meloidogyne hapla* | 15,358 | 37608 | 0 | 0 | 0 | 65 | 0 | 62 |
| Lloa | *Loa loa* | 15,825 | 174388 | 0 | 0 | 0 | 134 | 0 | 0 |
| Gpal | *Globodera pallida* | 18,139 | 121687 | 0 | 0 | 0 | 70 | 3 | 92 |
| Lsig | *Litomosoides sigmodontis* | 20,478 | 45863 | 0 | 0 | 0 | 3 | 0 | 0 |
| Acas | *Acanthamoeba castellanii* | 30,713 | 564894 | 29 | 4 | 21 | 23 | 58 | 46 |
| Nve | *Nematostella vectensis* | 33,008 | 472588 | 495 | 42 | 177 | 908 | 24 | 1099 |
| Dpul | *Daphnia pulex* | 38,001 | 642089 | 301 | 58 | 183 | 1012 | 455 | 1082 |
| Crem | *Caenorhabditis remanei* | 39,630 | 435512 | 16 | 3 | 11 | 149 | 0 | 88 |
| Tmur | *Trichuris muris* | 50,312 | 400602 | 0 | 0 | 0 | 520 | 114 | 496 |
| Cbre | *Caenorhabditis brenneri* | 57,601 | 381961 | 0 | 0 | 0 | 136 | 0 | 78 |
| Lgi | *Lottia gigantea* | 80,338 | 1870055 | 333 | 17 | 155 | 548 | 19 | 521 |
| Pred | *Panagrellus redivivus* | 97,659 | 270080 | 5 | 0 | 0 | 25 | 0 | 26 |
| C11sp | *Caenorhabditis* sp11 | 119,280 | 20921866 | 0 | 0 | 0 | 27 | 0 | 14 |
| Acal | *Aplysia californica* | 214,107 | 917541 | 64 | 9 | 33 | 202 | 26 | 538 |
| Alyr | *Arabidopsis lyrata* | 297,364 | 24464547 | 0 | 0 | 0 | 1209 | 1607 | 1454 |
| Vcar | *Volvox carteri* | 302,219 | 2599759 | 182 | 35 | 111 | 515 | 435 | 618 |
| Srat | *Strongyloides ratti* | 1,047,729 | 4921549 | 0 | 0 | 0 | 18 | 0 | 40 |
| briC | *Caenorhabditis briggsae* | 9,034,972 | 17485439 | 33 | 8 | 25 | 44 | 0 | 35 |
| Cele | *Caenorhabditis elegans* | 14,326,629 | 17493829 | 1 | 0 | 1 | 28 | 2 | 11 |
| Ath | *Arabidopsis thaliana* | 17,095,393 | 23459830 | 0 | 0 | 0 | 252 | 374 | 433 |
| Nvi | *Nasonia vitripennis* | 59,454,730 | 48524378 | 101 | 9 | 47 | 760 | 510 | 1428 |
| | | Total | | 2539 | 207 | 963 | 11264 | 4102 | 12165 |

YR matches are shown, of which, the number of YREs that were classified based on their features and their phylogenetic position is indicated. In addition, the counts of RT hits from Bel, Copia and Gypsy LTR elements are indicated for each species. Matches were found in all the species in at least one of the PSITBLASTN searches. The number of matches found in each species seems to be detached from the mean contig length or contig length at N50 in the species' genome assembly.

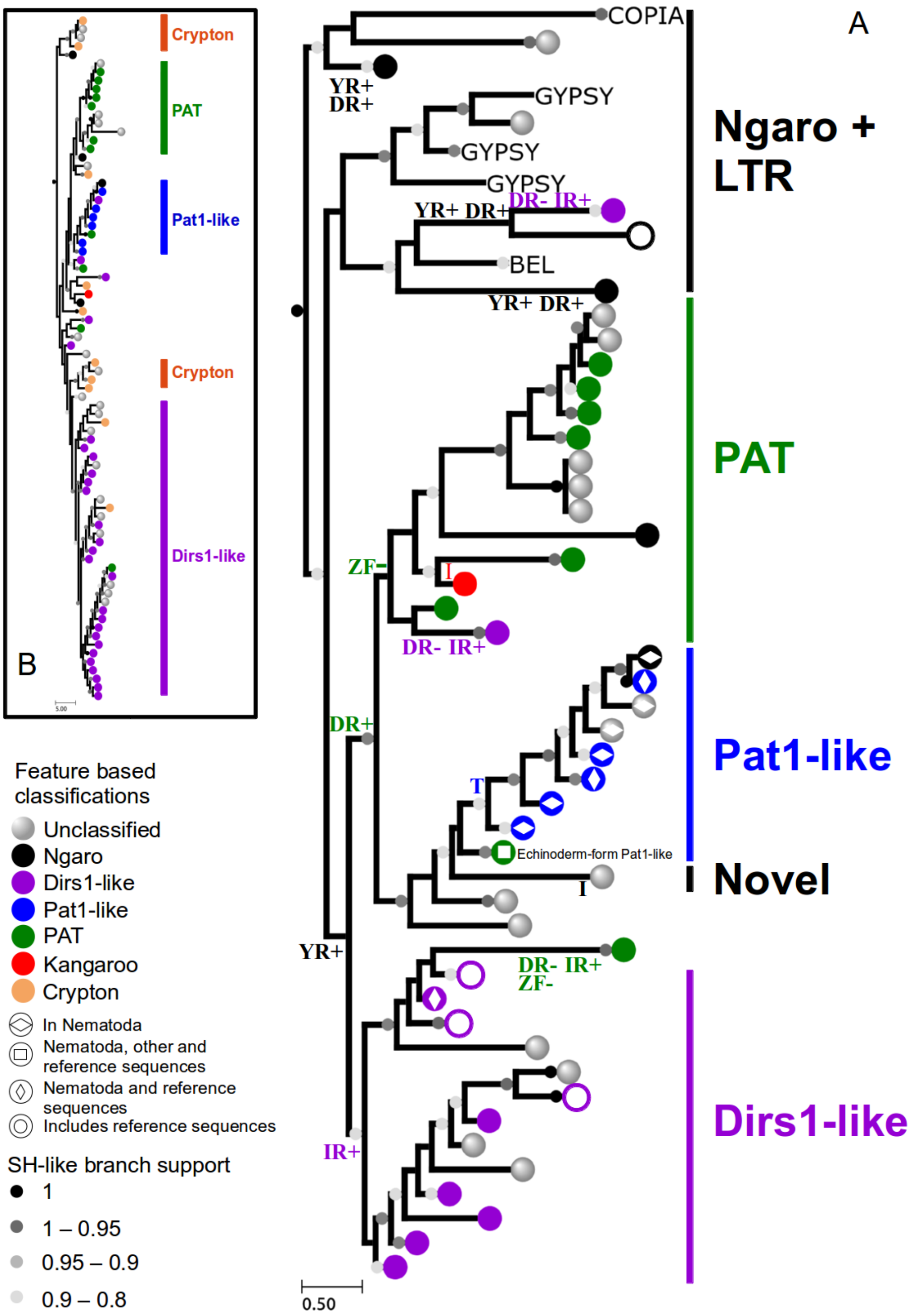

**Fig. 2: The phylogenetic relationships among YREs recovered from Nematoda and outgroup species**
The phylogeny of the YREs was derived from analyses of the RT domains (A) and the YR domains (B). Character state changes of diagnostic YRE features are indicated as follows: YR: tyrosine recombinase domains; DR: split direct repeats; IR: inverted repeats; I: inversion of the YR domain; T: translocation of the internal repeats; zf: zinc finger in the Gag protein. sh - like branch supports are indicated at the base of nodes. Feature based classification, and the inclusion of reference sequences is indicated on each leaf. Where the leaves have a branch support symbol, these leaves are in fact collapsed clades

*Partition homogeneity test*

Reciprocal AU-tests were conducted to test the phylogenetic homogeneity of the YR and RT domains, using datasets with identical element sampling. All the tests rejected the homogeneity of the two partitions, suggesting either a real difference in the phylogenetic history of the two markers, or low phylogenetic signal in one or both of the markers. Because the RT domain demonstrated a stronger phylogenetic signal, according to the sh-like node support values, we based our inference of the phylogenetic relationships between the different YRE lineages on phylogenetic analysis of the RT domains from complete YREs (Supplementary Methods).

**Phylogenetics and distribution of YREs in the studied genomes**

*Dirs1-like* elements

More than half of the recovered YREs were phylogenetically classified as *Dirs1*-like (504 elements). *Dirs1*-like elements were recovered as one major lineage and two or more additional minor lineages in the RT (fig. 2A) and YR (fig. 2B) trees, respectively. One of the minor lineages clustered among PAT elements in both the YR and RT trees. The major linage was paraphyletic (with respect to element classification by structural features; fig. 1) in both analyses and included a PAT group, which appeared to be misplaced in the RT tree (fig. 2A) due to its long branch.

Among the outgroup taxa, *Dirs1*-like elements were found in *Acanthamoeba castellanii*, Cnidaria, Mollusca and Arthropoda (fig. 3). In contrast to previous reports, *Dirs1*-like elements were also found in Nematoda. *Enoplus brevis* (Enoplida; Enoplia) and *Romanomermis culicivorax* (Mermitihida; Dorylaimia) had several *Dirs1*-like elements each (7 and 68, respectively). *E. brevis* elements were truncated and clustered with complete *Dirs1*-like elements from the arthropod *Daphnia pulex* (sh-like support of 0.96, fig. S2). The absence of intact elements in *E. brevis* is likely to be because of the short average contig length (447 bp) of this assembly. *R. culicivorax Dirs1*-like elements included five complete elements, which were most closely related to elements from the tardigrade *Hypsibius dujardini* (Parachela; Eutardigrada) (sh-like support > 0.95, fig. S2).

In Chromadoria, a single partial *Dirs1*-like element was found in *P. pacificus.* It clustered with complete *Dirs1*-like elements from *R. culicivorax* (sh-like support > 0.95, fig. S2). It had a long branch and no significant matches in the BLAST database and thus is marginal in terms of affirming YR ancestry. All the *Dirs1* instances found in Nematoda belong to the major *Dirs1*-like lineage (fig. 2).

*PAT elements*

A paraphyletic clade of PAT elements, including *Pat1*, *Kangaroo*, one novel form (fig. 1G) and PAT elements, which were not further classified, was recovered in the RT tree (fig. 2A). Its paraphyly was due to a single minor lineage of *Dirs1*-like elements, which clustered with the PAT lineages in both the RT and YR trees, and a single Ngaro lineage, which might be misplaced, considering its long branch. An additional PAT group clustered inside the *Dirs1* major linage. The *Pat1*-like lineage comprised 142 Nematoda-form *Pat1* elements (fig. 1C) and 27 echinoderm-form elements (fig. 1D). These 27 elements were classified as *Pat1*-like due to the presence of a zinc-finger motif in the Gag sequence of some of them, in addition to their phylogenetic position. They clustered together with the echinoderm-form *Pat1*-like sequence from Retrobase (SpPat1). The *Pat1*-like elements of both forms (fig. 1C and 1D) formed a monophyletic clade in the RT tree (fig. 2A). In this clade, the echinoderm-form elements were early diverging. In the reduced YR tree (fig. 2B), the two forms were recovered as separate lineages.

*Kangaroo* elements from the alga *V. carteri* (24 elements) were represented by a single lineage within the PAT clade (fig. 2A). A PAT element in the arthropod *D. pulex* was represented by four full and 6 truncated instances, clustered as a sister clade of the *Pat1*-like elements (labelled "novel", fig. 2A). It was similar to PAT elements in structure, though possessing an inverted YR domain (fig. 1G). Unlike *Kangaroo* elements, which also have inverted YR domains, the novel element had internal repeats upstream to the 3' terminal repeat and not between the MT and YR domains. The remaining unclassified PAT elements clustered paraphyletically in the RT tree (fig. 2A). However, they clustered into three different lineages in the reduced YR tree (fig. 2B).

Echinoderm-form *Pat1*-like elements (fig. 1D) were found in the dorylaimid nematode *R. culicivorax,* the mollusc *Lottia gigantiea,* the arthropod *Nematostella vecetensis* and the alga *V. carteri* (fig. 3). The *L. gigantiea* and *N. vecetensis Pat1*-like elements are most likely the same as the PAT elements reported in Piednoël et al. (2011). PAT elements lacking a Gag protein with a zinc-finger motif were found only outside Metazoa. Lacking a zinc-finger, these PAT elements could be considered to be *Toc3*-like (fig. 1F). However, many are partial elements from which Gag was not recovered. Thus, the precise identity of most PAT elements could not be determined.

The Nematoda-form *Pat1*-like elements (fig. 1C) were found in the nematode classes Dorylaimia and Chromadoria. In Chromadoria they were only detected in Rhabditomorpha and Diplogasteromorpha. The absence of *Pat1*-like elements from 23 out of the 29 sampled rhabditid species is surprising. Poor assembly quality cannot serve as the only explanation for this finding as several of the genomes lacking identified elements had good average contig length (table 1). The absence *Pat1* elements from *P. redivivus* was also unexpected, since this species is known to possess several *Pat1*-like elements (de Chastonay et al., 1992) and a reciprocal blast approach was taken to confirm this finding. The *P. redivivus* genome assembly was queried using BLAST with the first *Pat1* sequence which was originally described in *P. redivivus* (Genbank accession X60774). Twelve significant matches were found. For confirmation, these fragments were then used as queries to search the online NCBI BLAST database with default settings, detecting the original *Pat1* sequence (X60774) as a single hit. Since the matches were *Pat1* fragments that did not contain the YR ORF, they had not been recovered by our pipeline, and this lack of complete *Pat1* elements was likely due to incomplete assembly.

*Non DIRS YREs*

In the species surveyed we identified only a single Crypton element, in *Nematostella*, and this element has already been recorded in Repbase (locus Crypton-1_NV). An additional Crypton match in *Nasonia* was closely related to a previously identified element from oomycetes (locus CryptonF-6_PI in Repbase) and is a likely contamination. Using more lenient parameters, permitting larger clades with lower sh-like support to be included, increased the count of Crypton-like elements. However, this resulted in clades with simultaneous conflicting classifications. In addition, we identified three major lineages of Ngaro elements, including 182 instances that clustered with LTR elements. These lineages included the Ngaro reference sequences. An additional minor lineage, from *Caenorhabditis briggsae*, clustered closely with *Pat1*-like elements from the same species and showed minimal sequence divergence from them (fig. 2A). We suggest that this Ngaro lineage was a derived species-specific form of *Pat1* element that has lost its MT domain. Unlike Crypton elements, Ngaro elements were found in most of the animal phyla examined (fig. 3). Ngaro were abundant in the cnidarian *Nematostella vectensis* (114 instances) and in the mollusc *L. gigantea* (53 instances). However, within Ecdysozoa, Ngaro counts were lower and ranged between 2 in the nematode *E. brevis and 14* in *H. dujardini.*

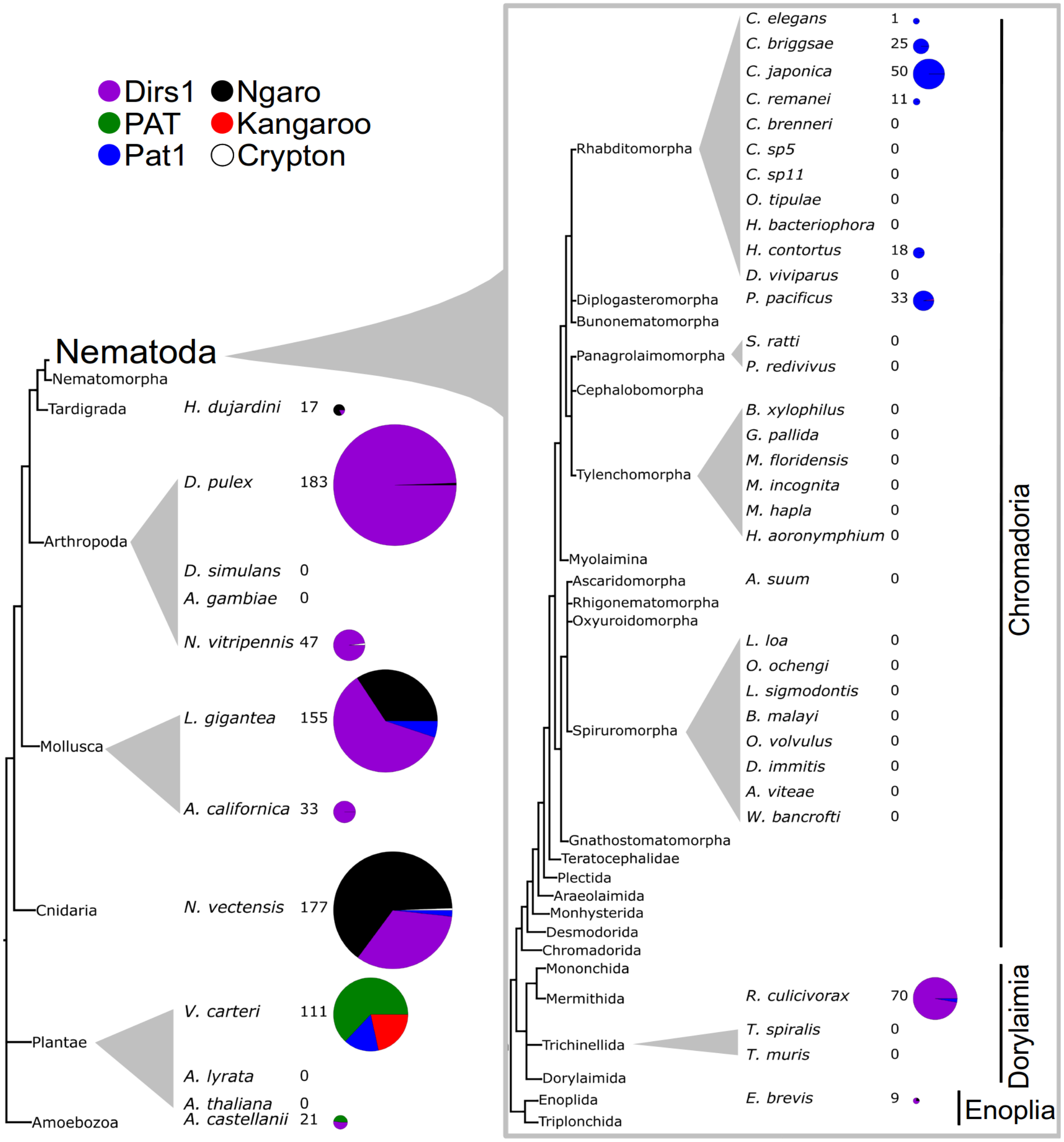

**Fig. 3: The distribution of YREs among Nematoda and outgroup species**
The phylogenetic tree of Nematoda is based on De Ley and Blaxter (2002) and Kiontke et al. (2013). Element types are colour coded. The phylogenetically classified YRE matches in each species are indicated. Pie-charts represent the proportion of each element type with their radii proportional to the number of phylogenetically classified YRE matches.

**The evolution of YRE features**

Based on the RT phylogeny, one of the possible most parsimonious scenarios for feature evolution is annotated in fig. 2. Under this hypothesis, the loss of the MT domain, the inversion of the YR domain, the formation of split direct repeats and of inverted repeats, and the loss of a zinc-finger motif have each occurred more than once, independently, and both split direct repeats and inverted repeats must have been formed through multiple sequential inversions. Any other possible scenario would require that several YRE features have evolved in parallel. In addition, any possible scenario would be inconsistent with single step character changes between element types: in figure S3 we hypothesized a scenario in which the different YRE retrotransposons were created only by single character changes in preexisting element types. This scenario is not supported by the phylogenetic analysis.

## Discussion

**Taxonomic representation in the study of TE**

The distribution of transposable elements has been hypothesized to depend on a number of factors; with mating system, ploidy, zygosity, ecology and gene flow all potentially influencing the TE load and diversity in an organism, in addition to the constraints of its phylogenetic history (Charlesworth and Charlesworth 1995; Wright et al. 2001; Wright and Finnegan 2001; Dolgin and Charlesworth 2006; Kawakami et al. 2010; Carr et al. 2012; Eads et al. 2012). Even within species, strains and populations can differ markedly in TE abundance (Collins et al. 1987). Therefore, when studying the distribution of TEs, it is unlikely that one would identify a single or a few species that would accurately represent a whole phylum, especially a phylum as species rich and diverse as Nematoda.

Piednoël et al. (2011) surveyed *Dirs1*-like YREs in a wide range of eukaryotes in order to understand the distribution of this element. Although 274 genome assemblies were analysed, only two nematode genomes were available to them, and these were from two closely related rhabditid superorders, Rhabditomorpha (*C. elegans*) and Diplogasteromorpha (*P. pacificus*). Neither species contained *Dirs1*-like sequences, leading to the conclusion that these elements were absent from nematodes as a whole. In this study, however, thanks to the wider taxonomic representation that is now available, we have identified *Dirs1*-like sequences in at least two out of the three nematode subclasses.

In addition, since many of the assemblies we screened were drafts and thus highly fragmented representations of the original genomes (the shortest average contig length was 411 bp in *H. aoronymphium*), we employed a search strategy that did not require the presence of complete YRE sequences, which may be as long as 6,000 bp (Piednoël et al. 2011). This approach, together with the classification of complete elements based on their structure, and the phylogenetic analysis of both complete and truncated elements, allowed us to recover and classify about 700 truncated YREs. To illustrate the power of this approach, while *Enoplus brevis* had an average contig length of 477 bp, we recovered nine elements that were classified based on their phylogenetic relationship with reference sequences, and which would have otherwise been missed. These results emphasize the importance of dense taxonomic sampling and of the inclusion of truncated elements in surveys of element diversity and distribution. Still, the failure to identify the expected *Pat1* elements in *P. redivivus* illustrates that the quantification and identification of TEs cannot be complete while focusing solely on protein domains and genome assemblies.

**YRE content in Nematoda has undergone a shift**

Based on our findings, Nematoda has undergone a substantial change in the composition and numbers of YREs (see fig. 3). The YRE content of the enoplid and dorylaimid species examined was more similar to that of outgroup taxa in *Dirs1* proportions than the YRE content of the rhabditid species. Indeed, *Dirs1*-like elements, relatively abundant in some outgroups, were found in *E. brevis* and *R. culicivorax* but were sparse in Rhabditida. The only potential *Dirs1*-like element found in Rhabditida was probably misclassified or a result of contamination, and *Dirs1*-like elements may be absent from Rhabditida altogether. In addition, the echinoderm-form *Pat1*-like element is found in *R. culicivorax* but not in other nematodes. It will be very informative to sample species from additional chromadorid superorders to identify the mode and tempo of this loss.

**The evolution of PAT elements**

The known distribution of the *Pat1* group of elements in Metazoa has been puzzling. *Pat1* elements were previously found only in Nematoda (Ecdysozoa)(de Chastonay et al. 1992; Poulter and Goodwin 2005; Piednoël et al. 2011) and Echinodermata (Deuterostomia) and the elements from these phyla have distinctly different feature organisations (de Chastonay et al. 1992; Poulter and Goodwin 2005; Piednoël et al. 2011) (Goodwin and Poulter 2004). The *Pat1* elements from these phyla have distinctly different feature organisation. Piednoël et al. (2011) were unable to classify the PAT elements from Cnidaria and Mollusca as *Pat1*. Consequently, the known distribution of the *Pat1* group of elements in Metazoa was puzzling. Here, through the phylogenetic classification of truncated elements, we identified the PAT elements in Mollusca and Cnidaria as *Pat1*-like, suggesting that these elements, though rare in general, are found in all three branches of Bilateria, and in non-bilaterian Metazoa. Surprisingly, the *Pat1*-like element that was found in the nematode *R. culicivorax* has the echinoderm-form structural arrangement rather than the nematode-form of *Pat1*. In addition, *Pat1*-like elements from Nematoda and from Echinodermata form sister clades in the RT tree (fig. 2A). Thus, the nematode-form *Pat1*-like element is not an isolated element with an unknown origin, but rather a taxon specific clade of a more widespread *Pat1* element family, and we suggest that there exists a greater diversity of these elements yet to be discovered by completed genome projects.

**Homoplasy in YRE structural features and the need for phylogenetics**

YREs have been suggested to have emerged from a composite ancestor combining an LTR element with a Crypton, as both Cryptons and LTRs are considered to be more ancient than YREs based on their distribution (Jurka et al. 2007). It is not clear, however, whether a single or several independent events of recombination are at the base of YRE retroelements. Our results support at least two origins for YRE retroposons: at least one for Ngaro elements and another for DIRS elements. As a consequence, split direct repeats must have evolved more than once, independently, resulting in homoplasious similarity. This result is in accordance with the phylogenetic tree presented in Goodwin and Poulter (2004). While Goodwin and Poulter (2004) found that PAT and *Dirs1*-like elements form a single clade each, we observed a paraphyletic, or possibly polyphyletic *Dirs1* group. Since this was observed in both the RT and YR trees (fig. 2), this result could either mean that PAT elements evolved from *Dirs1* or that a *Dirs1*-like element evolved twice independently. It is worth noting that the formation of inverted repeats from split direct repeats is a complex process that would require some intermediate forms. However, these forms are not observed, possibly due their inviability.

Another homoplasious similarity between

polyphyletic element lineages was observed in Ngaro and a derived lineage of *Pat1*-like elements in *C. briggsae*, both lacking a MT domain. In addition, a derived PAT element in *D. pulex* had homoplasious similarity to *Kangaroo* from *V. carteri*, both having an inverted YR domain. Also, we infer that the loss of a zinc-finger motif from the Gag protein must have occurred independently multiple times. Taking these observations together, homoplasy in element features is a strong theme in the evolution of YREs. This strongly suggests that it is impractical to use structural characteristics as the sole descriptors for element classification, and incorporating an explicitly phylogenetic basis for classification would produce more biologically meaningful inferences.

**Conclusions**

In this study we utilised a large number of nematode genome assemblies to characterize the YRE content in Nematoda. We showed that the YRE content across the phylum is much more diverse than suggested by the analysis of a few model species. It was important to include truncated elements to fill the gaps in the extant diversity of both *Dirs1*-like and *Pat1*-like elements, both of which are more widely distributed than originally perceived. Our results strongly support a previous call (Seberg and Petersen, 2009) to classify transposable elements based on phylogenetic relationships rather than the features they contain or lack, thus conforming to a systematic approach to classification.

## Material and Methods

The analyses presented here have been made reproducible and extendible by using a single IPython notebook for all the stages. A static html file of the notebook is included as Supplementary Methods. The live IPython notebook is published on github (https://github.com/HullUni-bioinformatics/Szitenberg_et_al_2014) and, with the exception of genome assemblies, the repository includes all the input files. URLs to the genome assemblies are provided in table S1. All the analyses and figures presented here can be reproduced by downloading the assembly files and executing the IPython notebook cells in sequence while following the instructions included in the notebook. However, since the assembly versions that were used here may be inaccessible in the future, all the pipeline's outputs are also provided in the github repository.

**Taxon sampling**

Our nematode species sampling consisted of 34 genome assemblies belonging to ten orders and superorders (fig. 2). Most of the species (30) belong to the subclass Chromadoria, three to the subclass Dorylaimia and one to Enoplia. Five ecdysozoan species, including four arthropods and a single tardigrade, were selected as outgroup taxa. Non-ecdysozoan outgroup species included a cnidarian, two molluscs, an amoebozoan and three plants. The species and sources are listed in table S1.

In addition to genome assemblies, we also analysed the Repbase Crypton and DIRS datasets (Jurka et al. 2005), the Retrobase DIRS dataset (http://biocadmin.otago.ac.nz/fmi/xsl/retrobase/home.xsl), four *Pat1*-like elements from *P. pacificus* kindly provided by M. Piednoël, and the first *Pat1* sequence to have been described (Genbank accession X60774). These sources pooled together formed our reference dataset. We examined the validity of element classifications produced by the pipeline using these known elements and also for seeding query alignments.

**Homology search based YRE identification**

In order to find YREs in the assemblies we used a strategy modified from Piednoël et al. (2011). First, we searched for YR domains in each whole genome assembly. YR matches were extended by 10 kbp in each direction or to the contig end, whichever was encountered first. We then searched for RT and MT domains and direct and inverted repeats in the resulting sequences. This approach efficiently streamlined the homology searches while including only RT and MT domains that are likely to belong to YREs. The homology searches were conducted using PSITBLASTN (Altschul et al. 1997; Camacho et al. 2009) with an expected value threshold of 0.01. The query models for these searches were seeded with the alignments from Piednoël et al. (2011) and were extended by adding protein sequences from the reference dataset through PSIBLASTP search (Altschul et al. 1997; Camacho et al. 2009).

Direct and inverted repeats on the extended YR fragments were detected with the BLAST based program UGENE (Okonechnikov et al. 2012), with only identical repeats at least 20 bp long allowed. These values represent the minimal repeat sequence in the results of Piednoël et al. (2011). Each annotated fragment was subsequently programmatically given a preliminary classification based on its similarity to the structures illustrated in fig. 1.

**Zinc-finger motifs pattern matching**

Among PAT elements (fig. 1), only *Pat1* elements have zinc-finger motifs in their Gag sequence (Poulter and Goodwin 2005). Gag sequences from two *Pat1* elements were used to query the reference databases to produce a Gag sequence model using PSIBLASTP (Altschul et al. 1997; Camacho et al. 2009). The sequences that were eventually used to produce the model represented all the DIRS elements' diversity. PSITBLASTN (Altschul et al. 1997; Camacho et al. 2009) was used to recover Gag sequences from the YRE DNA sequences found in the previous stage, with an expected value threshold of 0.01. The Gag sequences detected were searched for the zinc-finger sequence patterns described by Poulter and Goodwin (2005) using a python script (see Supplementary Methods). The classification process was continued later on, using a phylogenetic approach, to account for partial and degraded elements as well as for complete ones.

**Phylogenetic reconstruction of YRE relationships**

For the inference of phylogenetic relationships among YRE clades we considered only YRE matches that had at least YR and RT domains as well as terminal repeats. The RT domain may have had a different history from that of the YR domain as YR and RT trees from the literature do not seem to be congruent (Jurka et al. 2007; Kojima and Jurka 2011). Therefore, a reciprocal AU-test for partition homogeneity was conducted in CONSEL 0.2 (Shimodaira and Hasegawa 2001), using a RT, YR and combined datasets with identical YRE representation. Since the results indicated incongruence between the partitions (see Results and Supplementary Methods), and since preliminary analysis revealed better sh-like support in the tree that was reconstructed from the RT dataset, the RT domain was chosen for the phylogenetic reconstruction of YRE relationships. Gypsy, Copia and BEL sequences from Repbase were added to the RT dataset prior to the analysis. The RT sequences were aligned with MAFFT 7 (Katoh et al. 2002; Katoh and Standley 2013) using default settings and then trimmed with TrimAl 1.2 (Capella-Gutiérrez et al. 2009) to remove positions with over 0.3 gap proportion. The tree was reconstructed using FastTree 2.1.7 (Price et al. 2010) with gamma distribution of among site rate variation and with the JTT matrix of substitution rates (see Supplementary Methods for the exact command line parameters used).

**Phylogenetic approach to YRE classification and quantification**

We chose a phylogenetic approach to element classification over genetic-distance clustering methods to better account for homoplasy in our sequence data. Similar methods to the ones above were used to reconstruct two additional phylogenetic trees for the purpose of classification and quantification. The first tree was reconstructed from a dataset including only YR sequences from complete RNA YREs as well as Crypton YR sequences. This tree was used to delineate element clades. Only clades with sh-like support of 0.7 or above were considered, if they did not have conflicting YRE features based classifications. YR domain hits from reference elements helped to confirm the identity of the element clades.

The second tree included all the YR domain hits from both complete and truncated or degraded elements as well as YR sequences from Crypton elements. This tree was used to identify the phylogenetic position of degraded and truncated elements relatively to complete elements and adjust their count accordingly, for each of the clades recovered in the previous tree. Only truncated or degraded elements that clustered with complete elements with sh-like support of 0.9 or above were considered. However, we have detached nodes with long branches from clades that included complete elements and had sh-like value < 0.95, to avoid artifactual groupings. The branch-length cutoff that was used for node removal due to a long branch was four times the median branch-length of that clade.

**Assessment of the reliabilty of YRE counts**

Given that the originating genome does in fact contain YRE elements, draft genome assemblies could be missing YRE elements for two reasons: The first is that by being incomplete they may stochastically miss some elements. The second reason arises from the assembly algorithms used, where highly similar elements may yield assembly graphs that the algorithm rejects as being too complex, or of too high coverage, to include in the reported assembly contigs. Since YREs often have a low copy number (Cappello et al. 1984; Kojima and Jurka 2011) the second artefact is less likely, but a record of absence may simply reflect assembly quality. However, LTR retrotransposons are not likely to be absent from eukaryotic genomes and an inability to detect LTR elements would suggest that the assembly is simply not of sufficient quality. Therefore, in each of the species studied, we performed three additional PSITBLASTN (Altschul et al. 1997; Camacho et al. 2009) searches for RT domains of Gypsy, Copia and BEL LTR retrotransposons. The query alignments were constructed in the same manner as described above and are available in the github repository.

## Acknowledgements

We thank colleagues for agreeing access to unpublished nematode and other genome sequences: Einhardt Schierenberg for *E. brevis*; John Jaenicke for *H. aoronymphium*; Aziz Aboobaker for *H. dujardini*; Simon Bbabayan, Benjamin Makepeace, Kenneth Pfarr and Judith Allen for *O. ochengi, L. sigmodontis* and *A. viteae*; Asher Cutter and Marie-Anne Felix for *O. tipulae* and *Caenorhabditis* sp. 5. ; Dr. John Spieth for *Caenorhabditis* sp. 11 and *Heterorhabditis bacteriophora*; Dr. Patrick Minx for *Anopheles gambiae.* The GenePool has core support from the NERC (award R8/H10/56) and MRC (G0900740).

AS, DL and MB are supported in part by NERC award NE/J011355/1. GK is supported by a BBSRC Research Studentship and an ORS award.

## Supplementary material

### Table S1: Source of genomic data

Abbreviation, taxonomy and genome assembly information of the species studied.

| Abbreviation | Species | Higher rank | Lower rank | Genome version | Link to genome data |
|---|---|---|---|---|---|
| Acal | Aplysia californica | Mollusca | Gastropoda | 0 | http://www.broadinstitute.org/ftp/pub/assemblies/invertebrates/aplysia/AplCal3/A_californica_v0.assembly.fasta.gz |
| Acas | Acanthamoeba castellanii | Amoebozoa | Acanthamoebidae | 2010210 | ftp://ftp.hgsc.bcm.edu/AcastellaniNeff/Acas20100210/LinearScaffolds/Acas20100210.contigs.agp.linear.fa |
| Agam | Anopheles gambiae | Arthropoda | Insecta | 4.2 | http://genome.wustl.edu/pub/organism/Invertebrates/Anopheles_gambiae/assembly/Anopheles_gambiae_S-4.2/output/supercontigs.fa.gz |
| Alyr | Arabidopsis lyrata | Plantae | Angiosperms | 107 | ftp://ftp.jgi-psf.org/pub/compgen/phytozome/v9.0/Alyrata/assembly/Alyrata_107.fa.gz |
| Asuu | Ascaris suum | Nematoda | Ascaridomorpha | WS238 | ftp://ftp.wormbase.org/pub/wormbase/species/a_suum/sequence/genomic/a_suum.PRJNA62057.WS238.genomic.fa.gz |
| Ath | Arabidopsis thaliana | Plantae | Angiosperms | TAIR10 | ftp://ftp.arabidopsis.org/home/tair/Sequences/whole_chromosomes/ |
| Avit | Acanthocheilonema viteae | Nematoda | Spiruromorpha | 1 | http://acanthocheilonema.nematod.es |
| Bmal | Brugia malayi | Nematoda | Spiruromorpha | WS238 | ftp://ftp.wormbase.org/pub/wormbase/species/b_malayi/sequence/genomic/b_malayi.PRJNA10729.WS238.genomic.fa.gz |
| briC | Caenorhabditis briggsae | Nematoda | Rhabditomorpha | WS238 | ftp://ftp.wormbase.org/pub/wormbase/species/c_briggsae/sequence/genomic/c_briggsae.PRJNA10731.WS238.genomic.fa.gz |
| Bxyl | Bursaphelenchus xylophilus | Nematoda | Tylenchomorpha | 1.2 | ftp://ftp.sanger.ac.uk/pub/pathogens/Bursaphelenchus/xylophilus/Assembly-v1.2/BurXv1.2.supercontigs.fa.gz |
| C11sp | Caenorhabditis sp11 | Nematoda | Rhabditomorpha | 3.0.1 | http://genome.wustl.edu/pub/organism/Invertebrates/Caenorhabditis_sp11_JU1373/assembly/Caenorhabditis_sp11_JU1373-3.0.1/output/supercontigs.fa.gz |
| C5sp | Caenorhabditis sp. 5 | Nematoda | Rhabditomorpha | WS230 | ftp://ftp.wormbase.org/pub/wormbase/releases/WS230/species/c_sp5/c_sp5.WS230.genomic.fa.gz |
| Cang | Caenorhabditis angaria | Nematoda | Rhabditomorpha | WS238 | ftp://ftp.wormbase.org/pub/wormbase/species/c_angaria/sequence/genomic/c_angaria.PRJNA51225.WS238.genomic.fa.gz |
| Cbre | Caenorhabditis brenneri | Nematoda | Rhabditomorpha | WS238 | ftp://ftp.wormbase.org/pub/wormbase/species/c_brenneri/sequence/genomic/c_brenneri.PRJNA20035.WS238.genomic.fa.gz |
| Cele | Caenorhabditis elegans | Nematoda | Rhabditomorpha | WS235 | ftp://ftp.wormbase.org/pub/wormbase/species/c_elegans/sequence/genomic/c_elegans.WS235.genomic.fa.gz |

| Abbreviation | Species | Higher rank | Lower rank | Genome version | Link to genome data |
|---|---|---|---|---|---|
| Cjap | Caenorhabditis japonica | Nematoda | Rhabditomorpha | WS238 | ftp://ftp.wormbase.org/pub/wormbase/species/c_japonica/sequence/genomic/c_japonica.PRJNA12591.WS238.genomic.fa.gz |
| Crem | Caenorhabditis remanei | Nematoda | Rhabditomorpha | WS238 | ftp://ftp.wormbase.org/pub/wormbase/species/c_remanei/sequence/genomic/c_remanei.PRJNA53967.WS238.genomic.fa.gz |
| Dimm | Dirofilaria immitis | Nematoda | Spiruromorpha | 2.2 | http://dirofilaria.nematod.es |
| Dpul | Daphnia pulex | Arthropoda | Crustacea | 1 | http://genome.jgi.doe.gov/Dappu1/download/Daphnia_pulex.fasta.gz |
| Dsim | Drosophila simulans | Arthropoda | Insecta | 1.4 | ftp://ftp.flybase.net/genomes/Drosophila_simulans/dsim_r1.4_FB2012_03/fasta/dsim-all-chromosome-r1.4.fasta.gz |
| Dviv | Dictyocaulus viviparus | Nematoda | Rhabditomorpha | 1 | http://dictyocaulus.nematod.es |
| Ebre | Enoplus brevis | Nematoda | Enoplida | 1.1 | http://enoplus.nematod.es |
| Gpal | Globodera pallida | Nematoda | Tylenchomorpha | 30052012 | ftp://ftp.sanger.ac.uk/pub/pathogens/Globodera/pallida/Assembly/ARCHIVE/Gpal.genome.30052012.scaffolds.fa.gz |
| Haor | Howardula aoronymphium | Nematoda | Tylenchomorpha | 1 | http://nematodes.org/downloads/959nematodegenomes/blast/db/Howardula_aoronymphium_clc_1.fna |
| Hbac | Heterorhabditis bacteriophora | Nematoda | Rhabditomorpha | 1.2.1 | http://genome.wustl.edu/pub/organism/Invertebrates/Heterorhabditis_bacteriophora/assembly/Heterorhabditis_bacteriophora-1.2.1/output/contigs.fa.gz |
| Hcon | Haemonchus contortus | Nematoda | Rhabditomorpha | WS238 | ftp://ftp.wormbase.org/pub/wormbase/species/h_contortus/sequence/genomic/h_contortus.PRJEB506.WS238.genomic.fa.gz |
| Hduj | Hypsibius dujardini | Tardigrada | Eutardigrada | 2.3 | http://badger.bio.ed.ac.uk/H_dujardini/fileDownload/zip_download?fileName=nHd.2.3.abv500.fna |
| Lgi | Lottia gigantea | Mollusca | Gastropoda | 1 | ftp://ftp.jgi-psf.org/pub/JGI_data/Lottia_gigantea/v1.0/Lotgi1_assembly_scaffolds.fasta.gz |
| Lloa | Loa loa | Nematoda | Spiruromorpha | 3 | http://www.broadinstitute.org/annotation/genome/filarial_worms/download/?sp=EASupercontigsFasta&sp=SLoa_loa_V3&sp=S.zip |
| Lsig | Litomosoides sigmodontis | Nematoda | Spiruromorpha | 2.2 | http://litomosoides.nematod.es |
| Mflo | Meloidogyne floridensis | Nematoda | Tylenchomorpha | 1 | http://meloidogyne.nematod.es |
| Mhap | Meloidogyne hapla | Nematoda | Tylenchomorpha | WS238 | ftp://ftp.wormbase.org/pub/wormbase/species/m_hapla/sequence/genomic/m_hapla.PRJNA29083.WS238.genomic.fa.gz |
| Minc | Meloidogyne incognita | Nematoda | Tylenchomorpha | WS238 | ftp://ftp.wormbase.org/pub/wormbase/species/m_incognita/sequence/genomic/m_incognita.PRJEA28837.WS238.genomic.fa.gz |
| Nve | Nematostella vectensis | Cnidaria | Anthozoa | 1 | ftp://ftp.jgi-psf.org/pub/JGI_data/Nematostella_vectensis/v1.0/assembly/Nemve1.fasta.gz |
| Nvi | Nasonia vitripennis | Arthropoda | Insecta | 2 | ftp://ftp.hgsc.bcm.edu/Nvitripennis/fasta/Nvit_2.0/linearScaffolds/Nvit_2.0.linear.fa.gz |
| Ooch | Onchocerca ochengi | Nematoda | Spiruromorpha | 2 | http://onchocerca.nematod.es |

| Abbreviation | Species | Higher rank | Lower rank | Genome version | Link to genome data |
|---|---|---|---|---|---|
| Otip | Oscheius tipulae | Nematoda | Rhabditomorpha | 3.1 | http://nematodes.org/downloads/959nematodegenomes/blast/db/Oscheius_tipulae_clc3_1.fna |
| Ovol | Onchocerca volvulus | Nematoda | Spiruromorpha | 1 | http://www.broadinstitute.org/annotation/genome/filarial_worms/download/?sp=EASupercontigsFasta&sp=SOnchocerca_volvulus_V1&sp=S.zip |
| Ppac | Pristionchus pacificus | Nematoda | Diplogasteromorpha | WS238 | ftp://ftp.wormbase.org/pub/wormbase/species/p_pacificus/sequence/genomic/p_pacificus.PRJNA12644.WS238.genomic.fa.gz |
| Pred | Panagrellus redivivus | Nematoda | Panagrolaimomorpha | 1 | http://www.ncbi.nlm.nih.gov/nuccore?term=KB454917:KB455574[PACC] |
| Rcul | Romanomermis culicivorax | Nematoda | Mermithida | 1 | http://romanomermis.nematod.es |
| Srat | Strongyloides ratti | Nematoda | Panagrolaimomorpha | 4 | ftp://ftp.sanger.ac.uk/pub/pathogens/Strongyloides/ratti/version_4/Sratti_v4.genome.fa |
| Tmur | Trichuris muris | Nematoda | Trichinellida | 2b | ftp://ftp.sanger.ac.uk/pub/pathogens/Trichuris/muris/genome/version_2b/Tmuris_v2b.genome_scaffolds.fa |
| Tspi | Trichinella spiralis | Nematoda | Trichinellida | WS238 | ftp://ftp.wormbase.org/pub/wormbase/species/t_spiralis/sequence/genomic/t_spiralis.PRJNA12603.WS238.genomic.fa.gz |
| Vcar | Volvox carteri | Plantae | Chlorophyta | 9 | ftp://ftp.jgi-psf.org/pub/compgen/phytozome/v9.0/Vcarteri/assembly/Vcarteri_199.fa.gz |
| Wban | Wuchereria bancrofti | Nematoda | Spiruromorpha | 1 | http://www.broadinstitute.org/annotation/genome/filarial_worms/download/?sp=EASupercontigsFasta&sp=SWuchereria_bancrofti_V1&sp=S.zip |
| | Other reference data | | | | |
| PAT | Pristionchus pacificus Pat1-like | Reference | Reference | | Mathieu Piednoël personal communication |
| BAS | RepBase_Crypton | Reference | Reference | | http://www.girinst.org/protected/repbase_extract.php?division=&customdivision=&rank=&type=Crypton&autonomous=1&nonautonomous=&simple=1&format=EMBL&sa=Download |
| BAS | RepBase_DIRS | Reference | Reference | | http://www.girinst.org/protected/repbase_extract.php?division=&customdivision=&rank=&type=DIRS&autonomous=1&nonautonomous=&simple=1&format=EMBL&sa=Download |
| RET | Retrobase DIRS | Reference | Reference | | http://biocadmin.otago.ac.nz/fmi/xsl/retrobase/type1.xsl?-db=retrobase.fp7&-lay=AllFieldsLayout&-max=all&-sortfield.1=Subtype&-sortfield.2=Type&-sortfield.3=Family&Format=Method&-find |

**Fig. S1: Schematic description of the workflow utilized in this study**

A flow chart of the analysis steps described in the Material and Methods section, including the homology searches for YRE protein domains, the classification of YREs based on their features, the phylogenetic reconstruction of YRE relationships and their phylogenetic classification.

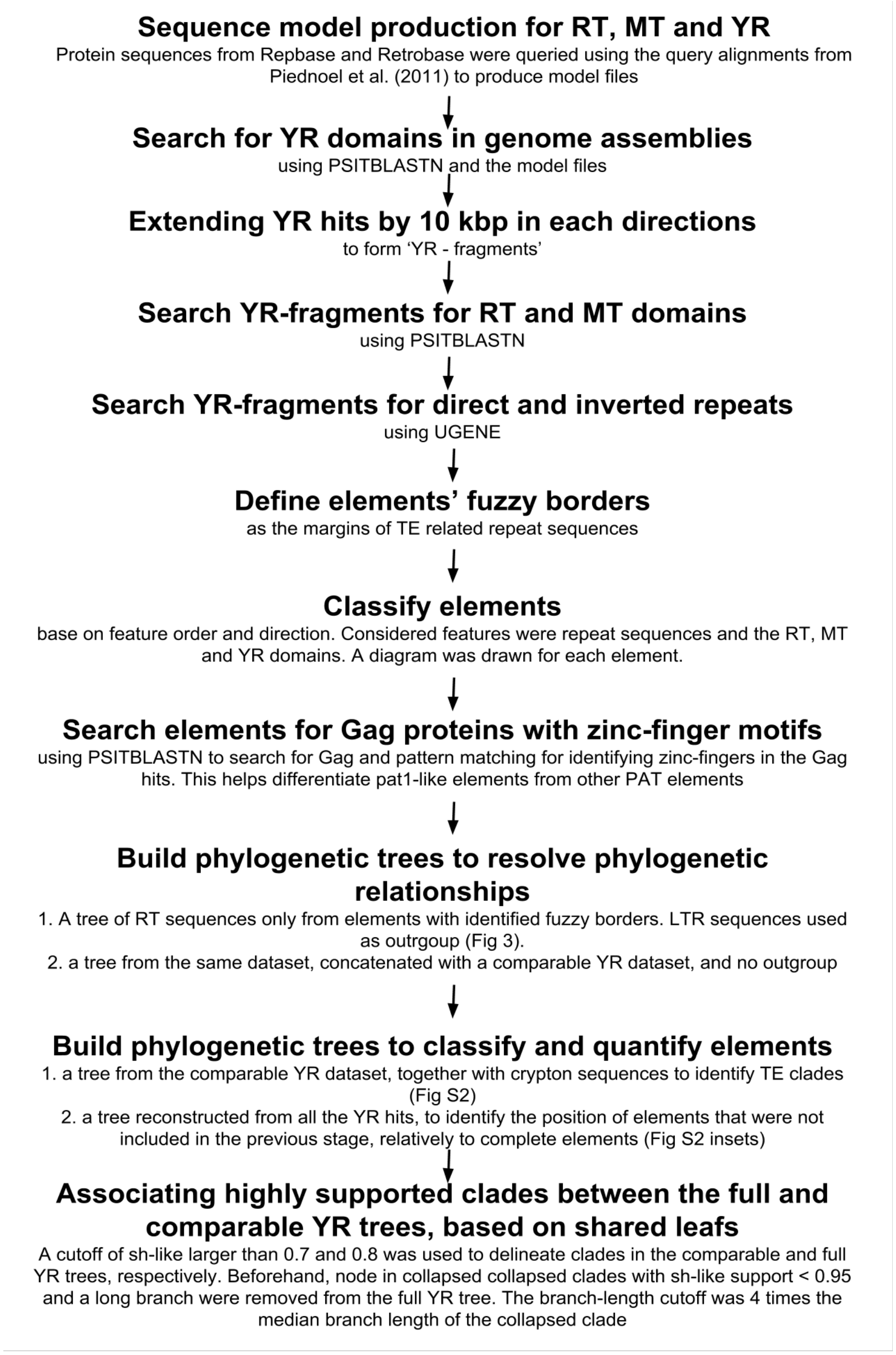

**Fig. S2: The phylogenetic classification of the recovered YREs**.

http://dx.doi.org/10.6084/m9.figshare.1009651

This phylogeny was reconstructed using only YR sequences from elements with defined borders (also available as feagure 2B), with a midpoint root (white background). Clades from the full YR tree (in grey) are presented next to reduced tree clades with which they share leaves. Large font black leaves are shared between the full and reduced YR trees. Large font green leaves are additional reference sequences. Small font leaves from the full tree (in grey) were added to the leaf count of the corresponding reduced tree clade. Only full tree clades with sh-like support > 0.9 were considered. Full tree clades that included long branches were removed if they had sh-like support < 0.95. The branch-length cutoff was four times the median branch-length of the clade. Leaf names include the species code (as in tables 1 and S1), a unique number and the feature based classification. The unique number is the start position of the YR domain on its contig. table.out files in the pipeline results folder (http://dx.doi.org/10.6084/m9.figshare.1004150) provide access to the complete element information using the species code and the unique number. The unique number provides access to the element's diagram in the same folder.

**Fig. S3: Hypothetical single step transitions between different YRE retrotransposon types.**

http://dx.doi.org/10.6084/m9.figshare.1009652

A flow chart depicting all the possible single step transitions between YRE retrotransposon types, using Ngaro as the ancestral form. Dirs1-like elements cannot be created from other element types in a single step. This scenario is not supported by the phylogenetic analysis (Fig. 2).

**Supplementary methods:**

http://dx.doi.org/10.6084/m9.figshare.1009650

The IPython notebook with which all the analyses related to this study were conducted is provided here as a static html file. It includes all the scripts along with detailed information. The executable IPython notebook is available in the github repository (http://dx.doi.org/10.6084/m9.figshare.1004150) along with the input and output files, except for the genome assemblies, which were very large. The genome assemblies can be accessed via links in Table S1 or in the iPython notebook.